\documentclass[aps,twocolumn,floatfix,showpacs]{revtex4}
\usepackage{graphicx,bm}
\begin{document}
\title{Variational approach for the two-dimensional trapped Bose Einstein 
condensate}
\author{Ludovic Pricoupenko}
\affiliation
{Laboratoire de Physique Th\'{e}orique des Liquides, 
Universit\'{e} Pierre et Marie Curie, 4 place Jussieu,
75252 Paris Cedex 05, France.}
\date{\today}
\begin{abstract}
We construct a many-body Gaussian variational approach for the two-dimensional 
trapped Bose gas in the condensate phase. Interaction between particles is 
modelized by a generalized pseudo-potential of zero range that allows recovering 
perturbative results in the ultra-dilute limit, while back action of non-condensate 
particles on the condensate part is taken into account for higher density. As an 
application, we derive the equation of state and solve stability problems 
encountered in similar mean-field formalisms for a single vortex configuration.
\end{abstract}
\pacs{03.75.Hh,05.30.Jp}
\maketitle

Both for fundamental reasons and for eventual future applications, ultra cold 
atomic gases trapped in reduced geometries has become a challenging field. 
Two-dimensional (2D) configurations are especially interesting for their various 
expected features including the interplay between the Bose Einstein and the 
Kosterlitz-Thouless transitions, the concept of tunable BEC \cite{Petrov1} 
and in the rotating case the possible realization of a gas with fractional 
statistics \cite{Cirac}. A quasi-2D BEC has been produced using polarized hydrogen atoms 
adsorbed on a helium film \cite{Safonov}, with clear evidence of the 
quasi-condensate phase. Several experimental groups 
\cite{MIT,Firenze,Innsbruck,Villetaneuse} are now interested in quasi-2D 
configurations using optical and magneto-optical traps which are expected to 
be more flexible for varying physical parameters such as the atomic density or 
the interaction strength. On the 
theoretical side, the general formalisms deeply differ from usual many body 
treatments in that they take into account both thermal phase fluctuations and 
the specific scattering properties 
\cite{Popov,Fisher,Andersen,Khawaja,Mora,Petrov2}.

	In this Letter we use the Gaussian variational approach of 
the many-body problem to study the condensate phase for the 2D BEC
\cite{Blaizot}. We show that describing interactions between atoms with the 
so-called $\Lambda$-potential \cite{Lambdapot} is a very simple way to 
construct a gapless and consistent theory.
In 3D configurations, this formalism has been already developed
\cite{Lambdapot} and allows to recover in a purely variational way the gapless 
HFB-G2 prescription \cite{Proukakis}. In the 2D regime, this way of taking 
into account the back action of non-condensed particles on the condensate 
field has not been yet implemented and brings new features. The Letter is 
composed of three parts. In the first part, we describe the modelization of 
the system and recall some basics of the two-body problem in 2D using the zero 
range approximation. In the second part, we develop the Gaussian variational 
approach (called also Hartree Fock Bogolubov or HFB) for the 2D trapped Bose 
gas and show the links with the HFB-G2 scheme. In the last 
part we apply the formalism in two distinct situations. We first calculate the 
Equation Of State (EOS) at zero temperature in the homogeneous limit. This 
first result generalizes the known perturbative EOS. Finally, as a test of the 
formalism we study the thermodynamical instability of a single vortex in a non 
rotating trap. We conclude that unlike unexpected results obtained in similar 
but non variational approaches 
\cite{Isoshima}, the system remains unstable at finite temperatures.
 
	We consider $N$ bosons of mass $m$ in a very anisotropic trap 
characterized by a high axial frequency $\omega_z$ and small longitudinal 
frequencies $\omega_x = \omega_y = \omega_L \ll \omega_z$. 
In the following, we suppose that the system is in a quasi two-dimensional 
geometry. This situation is achieved when the typical energy ($\epsilon$) of a 
two body scattering process is sufficiently low as compared to the axial 
frequency $|\epsilon| \ll \hbar \omega_z$.
In this regime, the system is somehow frozen along the tight direction in the 
ground state of the axial trapping potential and we can restrict our study to 
the longitudinal $xy$-plane. Before going into the approximate treatment of 
the many body problem, let us recall some basics of the two-body case in a 
planar trap ($\omega_L=0$). We note the spatial coordinates in the plane 
$(\vec{r}_1,\vec{r}_2)$, the relative and center of mass coordinates are 
respectively $\vec{r} = \vec{r}_1-\vec{r}_2$ and $\displaystyle 
\vec{R}=\frac{\vec{r}_1+\vec{r}_2}{2}$. In the zero range approach, the 
two-body wave function $\psi_2$ obeys the free Schr\"odinger equation for $r \neq 
0$, while the interaction is taken into account by imposing the contact 
condition~:
\begin{equation}
\psi_2(\vec{r}_1,\vec{r}_2) =  A(\vec{R}) \ln \left( \frac{r}{a_{2D}}
\right) \quad \mbox{for} \quad r\to0 \quad ,
\label{eq:contact}
\end{equation}
where $a_{2D}$ (the ``2D scattering length'') is obtained from the 
usual scattering length $a_{3D}$ and from the characteristic length of the 
trap $a_z$ using the relation \cite{Petrov2}:
\begin{equation} 
a_{2D} = 2.092 \, a_z \exp\left(-\sqrt{\frac{\pi}{2}} 
\frac{a_z}{a_{3D}}\right) .
\label{eq:a2D}
\end{equation}
Note that the short range part of the wave function is not described by 
relation (\ref{eq:contact}) \cite{Petrov2} which gives only the correct 
asymptotic behavior for $r \gg a_z$. However, the zero range approach is 
justified when the mean inter-particle distance is much larger than the axial 
width $a_z= \sqrt{\frac{\hbar}{m \omega_z}}$. In term of the 2D atomic density 
$n$, this gives $n a_z^2 \ll 1$. Another and equivalent way to implement the 
zero-range approach is to solve the Schr{\"o}dinger equation using the 
$\Lambda$-potential
\cite{Lambdapot} defined by:
\begin{equation}
\langle\vec{R},\vec{r}|V^{\Lambda}|\psi_2\rangle = 
g_\Lambda\,\delta^{(2)}(\vec{r}\,)\,
{\mathcal R}_\Lambda\!\!\left[\psi_2(\vec{R}-
\frac{\vec{r}}{2},\vec{R}+\frac{\vec{r}}{2})\right].
\label{eq:pseudo}
\end{equation}
In this potential, $g_\Lambda$ is the two-body T-matrix evaluated at energy $-
\hbar^2 \Lambda^2/m$~:
\begin{equation}
g_\Lambda = - \frac{2\pi\hbar^2}{m} \,\frac{1}{\ln(q \Lambda a_{2D})} \quad ,
\end{equation}
where $q = e^{\gamma}/2$, $\gamma$ is the Euler's constant and ${\mathcal 
R}_\Lambda$ 
is a regularizing operator: 
\begin{equation}
{\mathcal R}_\Lambda = \lim_{r \to 0} \left\{ 1-\ln(q \Lambda r)r \partial_{r} 
\right\} \quad.
\end{equation}
By construction, the observables are independent of the parameter 
$\Lambda$ which can be considered as a free field of the exact theory, 
function of the center of mass coordinates $\vec{R}$
\cite{Lambdapot}. However, we recall that some approximate treatments are 
$\Lambda$-dependent. In this type of formalism, $\Lambda$ appears as a new 
degree of freedom which is a way of improving the results. Taking for example 
the first order Born approximation in the two-body case, one can easily check that the 
exact scattering amplitude at energy $\epsilon$ is obtained using an uniform 
$\Lambda=-i \sqrt{m\epsilon}/\hbar$. In this case, $\Lambda$ is directly 
linked to the energy of the scattering process and has a straightforward 
physical meaning.

	Now, we implement the variational Gaussian approach in the case of a 2D 
condensate in the external potential $V_{\rm ext}=m \omega_L^2 R^2/2$
\cite{quasicondensate} using the $\Lambda$-potential (\ref{eq:pseudo}). 
In Ref.\cite{Lambdapot}, it has been already shown that this type of 
approximation is $\Lambda$-dependent. We explain in the following 
the criterion we use for a systematic determination of the field 
$\Lambda(\vec{R})$. In order to simplify the discussion, we break the $U(1)$ 
symmetry by splitting the atomic field $\hat{\psi}$ into a classical field 
$\Phi$ and a quantum fluctuation $\hat{\phi} = \hat{\psi} - \Phi$ \cite{BreakingU1}. 
The HFB scheme consists in choosing as a trial density operator the exponential 
of the most general quadratic form of the fields $\{\hat{\phi},\hat{\phi}^\dagger\}$. 
Diagonalization of the quadratic form is obtained using a Bogolubov 
transformation~:
\begin{equation}
\hat{\phi}(\vec{R}) = \sum_n \left\{ \hat{b}_n u_n(\vec{R}) 
	+ \hat{b}_n^\dagger v_n^*(\vec{R}) \right\} \quad ,
\end{equation}
where $\{\hat{b}_n,\hat{b}_n^\dagger\}$ are the usual creation and 
annihilation bosonic operators for a quasi-particle of quantum number $n$, 
while the amplitudes $\{ u_n,v_n \}$ verify the modal equations~:
\begin{eqnarray} 
&&\left[-\frac{\hbar^2}{2m} \Delta + V_{\rm ext} + \hbar \Sigma_{11} - \mu 
\right]\!\!
u_n + \hbar \Sigma_{12} v_n = \hbar \omega_n u_n \nonumber \\
 \label{eq:modes}
\\
&&\left[-\frac{\hbar^2}{2m} \Delta + V_{\rm ext} + \hbar \Sigma_{11}^* - \mu 
\right]\!\!
v_n + \hbar \Sigma_{12}^* u_n = - \hbar \omega_n v_n .
\nonumber 
\end{eqnarray}
From the variational principle \cite{Blaizot} we deduce the self-energies~:  
\begin{eqnarray}
&& \hbar\Sigma_{11}(\vec{R})=2 g_\Lambda \langle \hat{\psi}^{\dagger}(\vec{R})  
\hat{\psi}(\vec{R}) \rangle \label{eq:sigma11}\\
&& \hbar\Sigma_{12}(\vec{R})=g_\Lambda{\mathcal 
R}_{\Lambda}\!\!\left[\langle\hat{\psi}(\vec{R}-\frac{\vec{r}}{2}) 
\hat{\psi}(\vec{R}+\frac{\vec{r}}{2}) \rangle \right]
\label{eq:sigma12} \quad ,
\end{eqnarray}
and also a generalized Gross-Pitaevskii equation for the classical field
\begin{equation} 
\left[-\frac{\hbar^2}{2m} \Delta + V_{\rm ext} + 2 g_\Lambda \tilde{n} - \mu 
\right] \Phi
+ \hbar \Sigma_{12} \Phi^* = 0 
\label{eq:phi} \quad ,
\end{equation}
where $\tilde{n}$ is the non-condensed density. Let us insist on the fact that 
the modal equations (\ref{eq:modes}) do not coincide with the ones appearing 
in the linearized time-dependent treatment (the so-called RPA formalism; see 
for example Ref.\cite{Blaizot}). As a consequence, while the condensate mode associated to 
a phase change in the condensate wave function is a solution of the time 
dependent equations with a zero eigen-frequency, it is not in general solution 
of Eqs.(\ref{eq:modes}). In the homogeneous case, this leads to the presence of 
a gap in the spectrum defined through Eqs.(\ref{eq:modes}), and in the
inhomogeneous case, to the presence of an unphysical mode of non-zero
eigen-frequency. Unfortunately, the existence of a spurious energy
scale introduced by this mode prevents from 
a consistent description of the thermodynamical properties of the system. 
We show in the following that this problem can be solved without modifying the equations 
deduced from the variational principle. For that purpose, we recall that Eq.(\ref{eq:sigma11}) 
corresponds to a first order Born approximation so that unlike an exact approach, 
HFB is explicitly $\Lambda$-dependent. The key point is then to choose a 
specific realization of the field $\Lambda$, noted in the following 
$\Lambda^\star$, such that the condensate mode: 
\begin{equation}
u_0(\vec{R}) = \Phi(\vec{R}) \quad ; \quad  
v_0(\vec{R}) = -\Phi^*(\vec{R}) \quad ,
\end{equation}
solution of the time dependent equations, is also solution of Eqs.(\ref{eq:modes}) with a zero frequency.
This requirement leads to the condition 
$\hbar \Sigma_{12} = g_{\Lambda^{\star}} \Phi^2$ \cite{molecules}, so 
that at each point $\vec{R}$ in the condensate, $\Lambda^\star(\vec{R})$ is an 
implicit solution of the equation \cite{otherformalism}:
\begin{equation}
{\mathcal R}_{\Lambda^{\star}}\left[\langle \hat{\phi}(\vec{R}-
\frac{\vec{r}}{2})
\hat{\phi}(\vec{R}+\frac{\vec{r}}{2})\rangle \right] = 0 \quad .
\label{eq:zerogap}
\end{equation}
\begin{figure}[t]
\resizebox{8 cm}{!}{\includegraphics{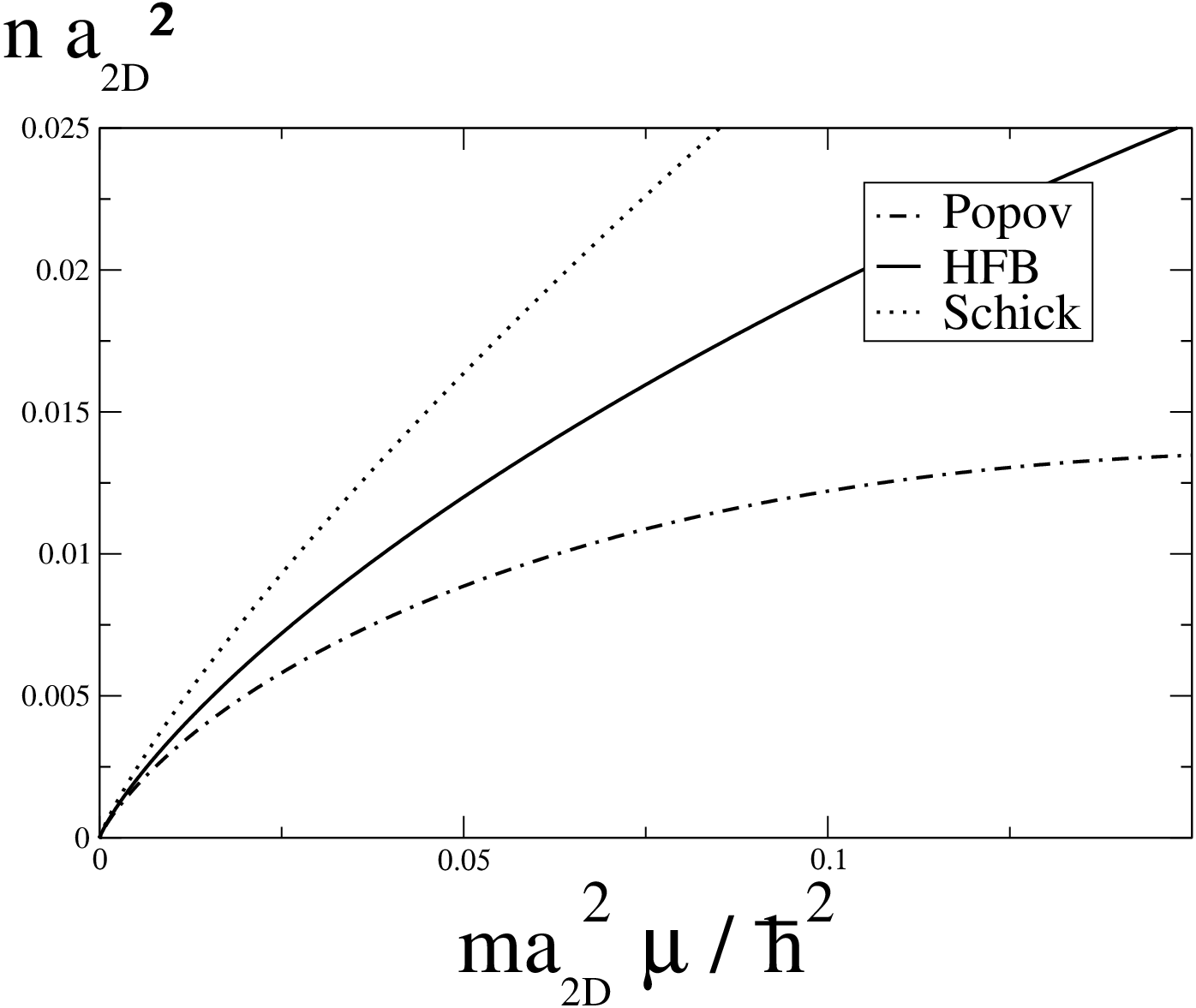}}
\caption{Equation Of State of the 2D homogeneous Bose gas. Dotted line: EOS 
deduced from Schick's formula
$\mu = 4 \pi \hbar^2 n/(m\ln(na_{2D}^2))$\cite{Schick}. Full line: full 
variational approach (HFB). Dashed dotted line: Perturbative EOS 
\cite{Popov,Mora}.}
\label{fig:eos}
\end{figure}
\begin{figure}[b]
\resizebox{8 cm}{!}{\includegraphics{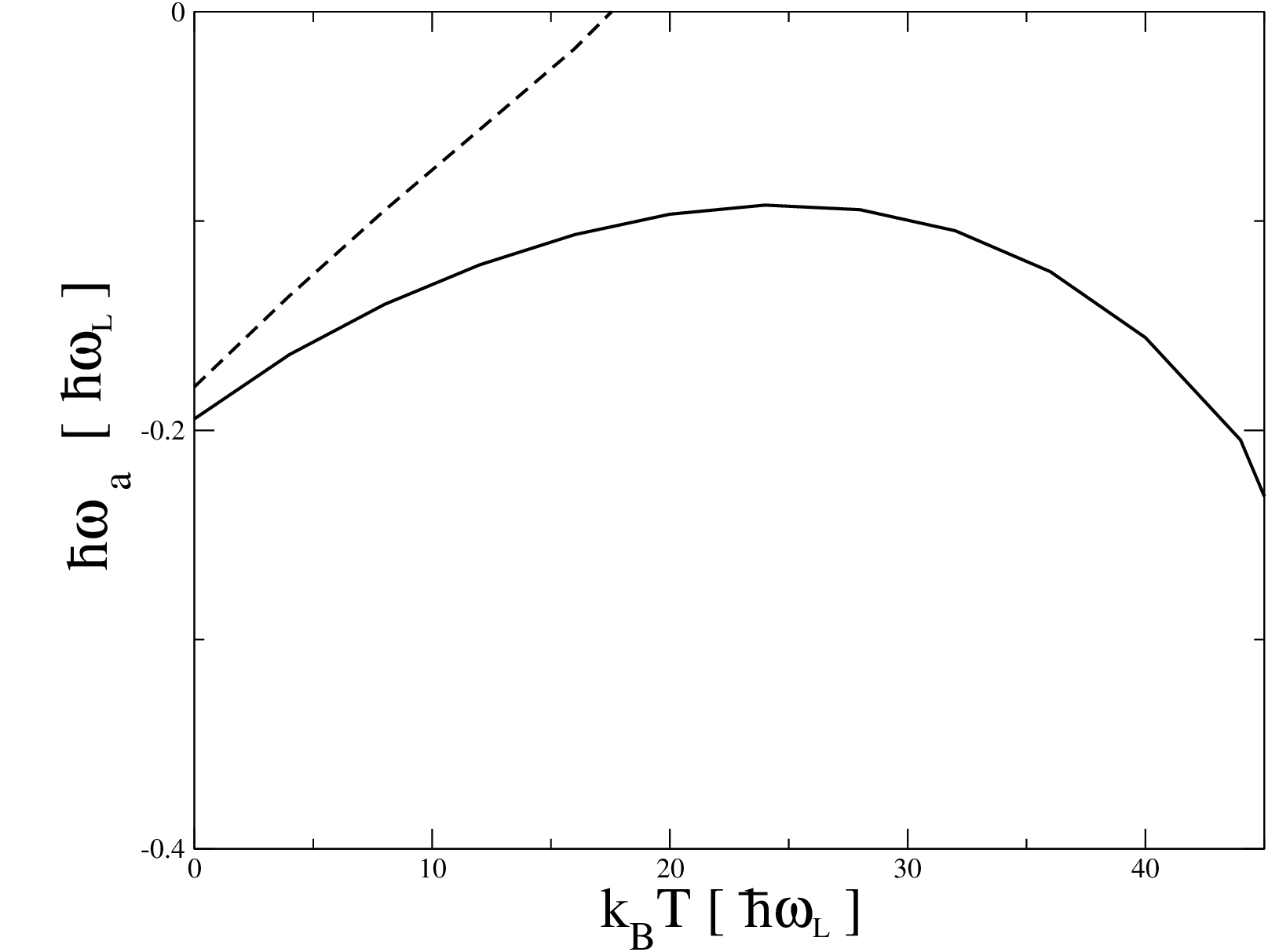}}
\caption{
Typical evolution of the anomalous mode energy of a single vortex in a non 
rotating trap as a function of the temperature. The presence of a negative 
energy mode in the spectrum leads to the so-called thermodynamical instability 
of the single vortex \cite{Rokhsar}. Dashed line: the coupling constant used 
in the gapless equations is arbitrarily evaluated using the two-body T-matrix 
at energy
$-2\mu_{\rm loc}$. Full line: full variational scheme; at finite temperature, 
the anomalous mode energy is not stabilized.}
\label{fig:instab}
\end{figure} 
\begin{figure}[b]
\resizebox{8 cm}{!}{\includegraphics{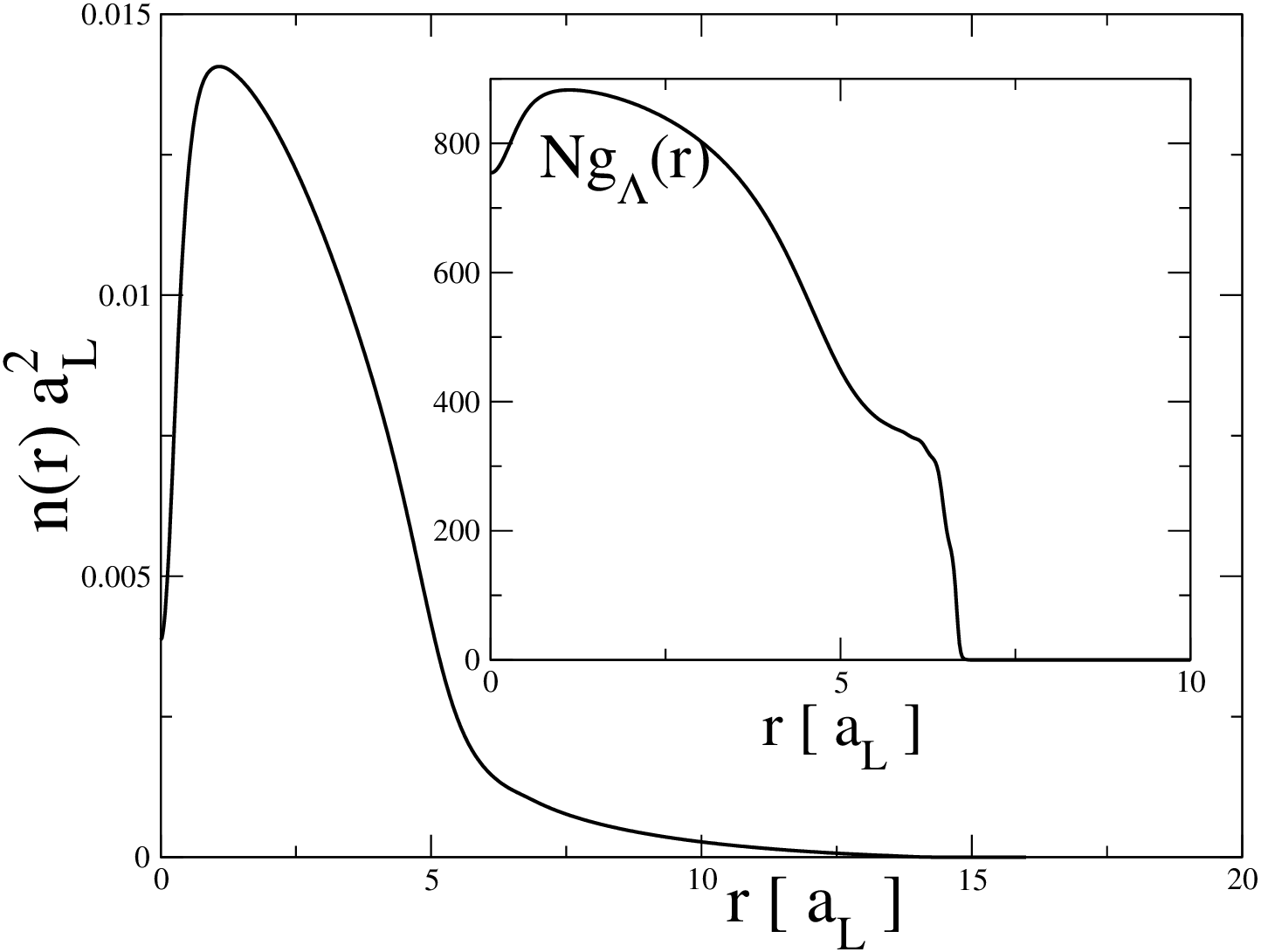}}
\caption{Typical density profile of a single vortex at finite temperature 
($k_BT=20\hbar\omega_L$), obtained using the full variational approach when 
the anomalous mode is not thermalized. Inset: mean-field coupling constant in 
harmonic oscillator units, obtained by solving Eq.(\ref{eq:zerogap}).}
\label{fig:vortex}
\end{figure}
In the limit $a_{3D} \ll a_z$ where interatomic collisions are 
weakly modified by the transverse confinement, Eq.(\ref{eq:zerogap}) leads 
at finite temperature to the HFB-G2 prescription \cite{Proukakis} introduced 
in the 3D configuration~:
\begin{equation}
	g_{\Lambda^{\star}}(T) \simeq  \frac{4 \pi \hbar^2 a_{3D}}{m a_z \sqrt{2 
\pi}} 
\left( 1 + \frac{\tilde{\kappa}_T}{\Phi^2} \right) \quad ,
\label{eq:glambdaT}
\end{equation}
where $\tilde{\kappa}_T(\vec{R})=\sum_n 2 g_n u_n(\vec{R}) v_n^*(\vec{R})$ is 
the thermal contribution of the pairing field and $g_n$ is the usual Bose 
occupation factor for the mode of energy $\hbar \omega_n$.

As a first application, we consider the homogeneous medium at zero 
temperature. From Eqs.(\ref{eq:modes},\ref{eq:phi},\ref{eq:zerogap}), we 
deduce a standard Bogolubov spectrum and a self-consistent equation for 
$\Lambda^\star$:
\begin{equation}
\frac{\hbar^2 \Lambda^{\star2}}{m} = g_{\Lambda^\star} |\Phi|^2 \quad ,
\label{eq:LDA}
\end{equation}
giving directly the energy at which the T-matrix has to be evaluated for 
having a gapless spectrum. From $\mu= g_{\Lambda^\star}(|\Phi|^2+ 2\tilde{n})$ 
and (\ref{eq:LDA}) we deduce the EOS shown in Fig.(\ref{fig:eos}).
In the ultra-dilute limit $na_{2D}^2\ll 1$, we recover the exact perturbative 
result \cite{Popov,Mora}:
\begin{equation}
n = \frac{m\mu}{4\pi \hbar^2}\ln\left(\frac{4 \hbar^2}{e^{(2\gamma+1)}m\mu 
a_{2D}^2} \right) \quad.
\label{eq:popov}
\end{equation}
For increasing values of the 2D atomic density, the predicted EOS deviates 
from Eq.(\ref{eq:popov}) (for a value of the 2D gas
parameter greater than $3\,10^{-5}$, the relative difference between
the two EOS is more than 5\%). Note that similarly to the 3D 
situation, there is an upper bound for the values of the density that can be 
studied \cite{Lambdapot}, here at $T=0$, $n<n_{\rm crit}=
\exp(-2\gamma)/\pi a_{2D}^2$. At $n=n_{\rm crit}$, $\mu \to \infty$ and
$\Phi\to0$, this gives a fundamental limit of applicability of the present 
formalism. In 2D traps, we have checked that at zero temperature, for static 
configurations, the full variational approach gives the same results as the 
LDA. In situations where the density varies rapidly, for example in presence 
of vortices, or also at finite temperature not only
Eqs.(\ref{eq:modes},\ref{eq:phi}) but also Eq.(\ref{eq:zerogap}) have to be 
solved numerically. We consider then as a last example, the case of a single 
vortex in a non rotating trap at finite temperature \cite{precisions}.
This configuration is especially interesting as a test of the present 
formalism. Indeed, a previous self-consistent but not variational approach 
\cite{Isoshima} has led to the conclusion that a vortex which is 
thermodynamically unstable at vanishing temperatures, could be stabilized at 
finite temperature. We recall that this instability is due to the existence of 
a core localized state having a  negative energy ($\hbar \omega_a$) when the trap 
rotation frequency is zero \cite{Rokhsar}. In 3D rotating BEC experiments, 
this type of instability which causes the spiraling of the vortex core out of 
the condensate has been observed \cite{Dalibard}, and we expect a similar 
behavior in 2D. We then consider an off-equilibrium situation where the trap 
is at rest and all the modes except the anomalous core localized state are 
thermalized. In a first step, we have checked that accordingly to 
Ref.\cite{Isoshima}, stabilization occurs when the coupling 
constant $g_\Lambda$ is arbitrarily set to a constant value in the ``gapless'' 
HFB equations. In a second step, we have used a more realistic prescription by 
assuming that the T-matrix entering the ``gapless'' equations is taken at 
energy $-\hbar^2 \Lambda^2/m=-2\mu_{\rm loc}$ ($\mu_{\rm loc}$ is the local 
chemical potential). As shown in Fig.\ref{fig:instab}, this non-variational 
approach predicts also an unphysical stabilization at sufficiently high 
temperature. On the contrary, using the full variational scheme where 
Eq.(\ref{eq:zerogap}) is solved numerically, the anomalous mode is never stabilized (see 
Fig.\ref{fig:instab}). In Fig.(\ref{fig:vortex}), we show a typical density profile and the 
associated mean-field coupling constant $g_{\Lambda^\star}$ obtained at finite 
temperature as a function of the distance from the core (we have used the 
notation $a_L=\sqrt{\hbar/m\omega_L}$). As explained in Ref.\cite{Isoshima}, 
the unexpected stabilization in the non-variational approach is linked 
to the mean-field term $g_\Lambda \tilde n$ in Eq.(\ref{eq:phi}) which acts 
as a pinning potential. In the formalism presented here, the low energy modes 
which have a very different structure with respect to the homogeneous case, 
crucially determine the behavior of the coupling constant $g_{\Lambda^{\star}}$ 
through Eq.(\ref{eq:zerogap}). As a consequence, the homogeneous result 
Eq.(\ref{eq:LDA}) cannot be applied: the lowest modes which become widely occupied 
with increasing temperatures, shape the structure of $g_{\Lambda^\star}$. For 
example in Fig.(\ref{fig:vortex}), the coupling constant dives at the origin 
(recall that the structure at the vicinity of the core plays a central role 
in the pinning process) and presents a buttress at the edge of the condensate 
due to the population of the surface modes.

	In conclusion, we have presented an approach for the 2D trapped BEC 
which takes into account the back action of the non-condensed fraction on 
the condensate field at finite temperature. This formalism includes known 
results on the 2D BEC and solves stability problems encountered in similar 
but non variational approaches.

{\bf Acknowledgments :} Y. Castin, C. Mora, M. Olshanii and H. Perrin are 
acknowledged for thorough discussions on the subject. LPTL is UMR 7600 of CNRS.


\begin{thebibliography}{99}
\bibitem{quasicondensate} A formalism describing a condensate phase in 2D may 
appears somewhat surprising. However, it is worth pointing out that for a 
large interval of temperatures and for typical values of the trap anisotropy 
parameter $\omega_z/\omega_L$, that can be reached in near future experiments 
(from $10^2$ to $10^3$), thermal phase fluctuations characteristic of the low 
dimensionality are likely to be quenched while the system keeps its 2D 
character. Taking as an example
a typical value of the chemical potential $\mu=20\hbar \omega_L$ for
$N=10^4$ atoms, even at the BEC transition temperature $k_B
T_C=\hbar\omega_L \sqrt{6N}/\pi$, the Thomas Fermi radius is smaller than the 
characteristic radius over which  phase fluctuations develop 
\cite{Petrov1}.
\bibitem{BreakingU1} HFB can be considered as a variant of the Girardeau-Arnowitz
approach \cite{Girardeau}. As a consequence, a generalization of HFB without 
breaking the U(1) symmetry is possible but goes beyond the scope of the 
present Letter.
\bibitem{molecules} The potential (\ref{eq:pseudo}) supports a bound state. As a 
consequence, HFB equations (\ref{eq:modes},\ref{eq:sigma11},\ref{eq:sigma12}) 
can lead to non trivial solutions: $\Phi=0$ whereas $\tilde n\neq0$ corresponding 
to a molecular condensate composed of dimers (see \cite{toy} for the 3D case). 
Condition (\ref{eq:zerogap}) allows avoiding this problem.
\bibitem{otherformalism} When $\Lambda$ is chosen using Eq.(\ref{eq:zerogap}), 
Eqs.(\ref{eq:modes},\ref{eq:phi}) are very similar to the one obtained in 
Refs.\cite{Andersen,Khawaja} which can describe also the quasi-condensate phase. 
However, comparison of the two formalisms makes sense only in the condensate 
phase. In this regime, while in Refs.\cite{Andersen} the T-matrix appearing in 
the mean-field equations is evaluated at energy  $-2 \mu_{\rm loc}$, where 
$\mu_{\rm loc}$ is the local chemical potential, in the present case the 
energy depends self-consistently on the structure and energy of the eigen-modes.
\bibitem{precisions} To be more precise, in the following calculations,
we have considered $N=10^4$ Cs atoms in a trap characterized by the 
atomic frequencies $\omega_L=2\pi\ 10$~Hz and $\omega_z=2\pi\ 10^3$~Hz.
In order to have a reasonable value for $\mu$ (on the order of $20\hbar 
\omega_L$), the 3D scattering length has been set arbitrarily to
$a_{3D}=35a_0$. Such a choice is made experimentally possible by use of a 
Feshbach resonance. In numerical simulations, the energy cut-off is $170 \hbar\omega_L$.
\end{thebibliography}
\end{document}